\documentclass[journal=jacsat,manuscript=communication]{achemso}
\setkeys{acs}{maxauthors = 0}
\usepackage{chemformula} % Formula subscripts using \ch{}
\usepackage[T1]{fontenc} % Use modern font encodings
\usepackage{amssymb}
\usepackage[symbol]{footmisc}

\author{Riaz Hussain}
\affiliation[Parma]{Dipartimento di Scienze Matematiche, Fisiche e Informatiche, Universit\`{a} di Parma, I-43124 Parma, Italy.}
\author{Giuseppe Allodi}
\affiliation[Parma]{Dipartimento di Scienze Matematiche, Fisiche e Informatiche, Universit\`{a} di Parma, I-43124 Parma, Italy.}
\author{Alessandro Chiesa}
\affiliation[Parma]{Dipartimento di Scienze Matematiche, Fisiche e Informatiche, Universit\`{a} di Parma, I-43124 Parma, Italy.}
\author{Elena Garlatti}
\affiliation[Parma]{Dipartimento di Scienze Matematiche, Fisiche e Informatiche, Universit\`{a} di Parma, I-43124 Parma, Italy.}
\alsoaffiliation[ISIS]{ISIS Facility, Rutherford Appleton Laboratory, OX11 0QX Didcot, United Kingdom.}
\author{Dmitri Mitcov}
\affiliation[Denmark]{Department of Chemistry, University of Copenhagen, DK-2100 Copenhagen, Denmark.}
\author{Andreas Konstantatos}
\affiliation[Denmark]{Department of Chemistry, University of Copenhagen, DK-2100 Copenhagen, Denmark.}
\author{Kasper S. Pedersen}
\affiliation[Denmark]{Department of Chemistry, University of Copenhagen, DK-2100 Copenhagen, Denmark.}
\alsoaffiliation[Denmark2]{Department of Chemistry, Technical University of Denmark, DK-2800, Kgs. Lyngby, Denmark.}
\author{Roberto De Renzi}
\affiliation[Parma]{Dipartimento di Scienze Matematiche, Fisiche e Informatiche, Universit\`{a} di Parma, I-43124 Parma, Italy.}
\author{Stergios Piligkos}
\affiliation[Denmark]{Department of Chemistry, University of Copenhagen, DK-2100 Copenhagen, Denmark.}
\author{Stefano Carretta}
\affiliation[Parma]{Dipartimento di Scienze Matematiche, Fisiche e Informatiche, Universit\`{a} di Parma, I-43124 Parma, Italy.}
\alsoaffiliation[INSTM]{UdR Parma, INSTM, I-43124 Parma, Italy.}
\email{stefano.carretta@unipr.it}

\title[Ybtrensal]
  {Coherent manipulation of a molecular Ln-based \\ nuclear qudit coupled to an electron qubit}

\abbreviations{IR,NMR,UV}

\begin{document}

\begin{tocentry}
	\centering
	\includegraphics[width=0.9\textwidth]{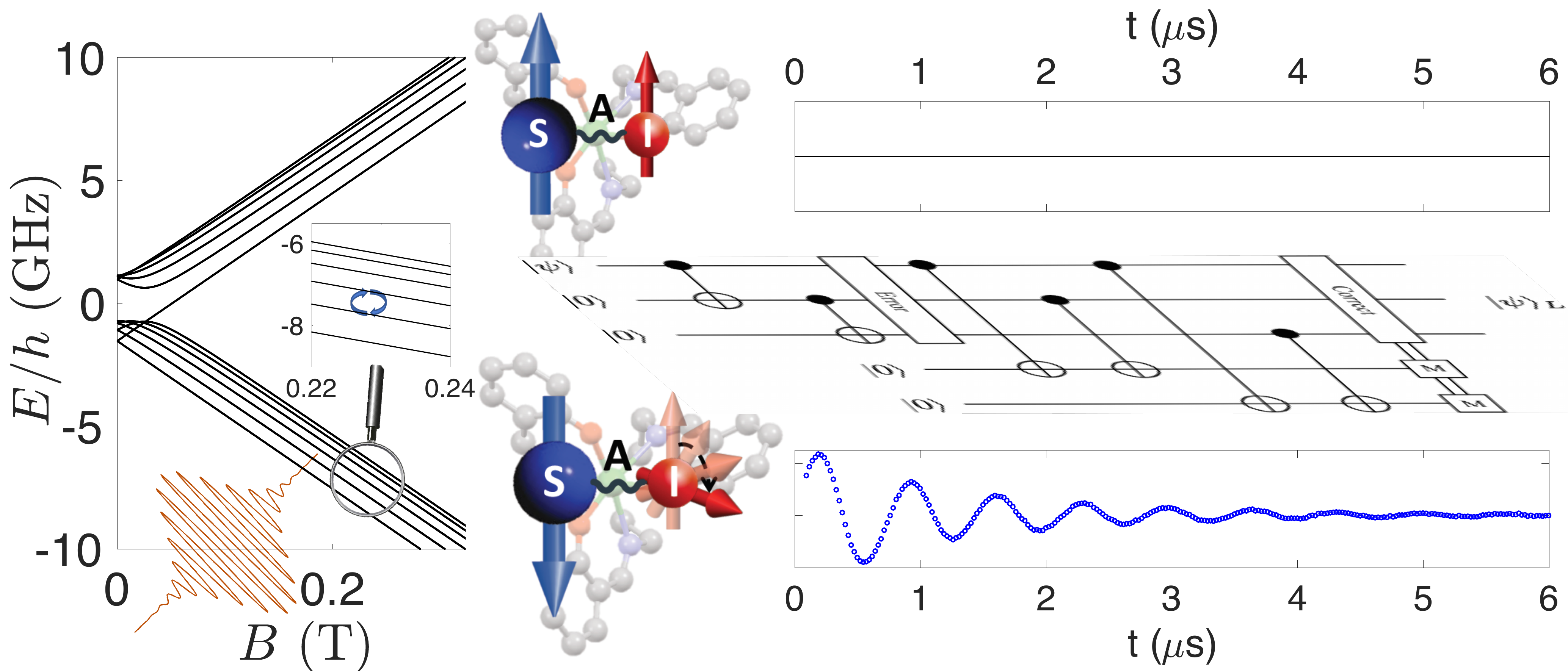}
\end{tocentry}

\begin{abstract}
We demonstrate that the [Yb(trensal)] molecule is a prototypical coupled electronic qubit-nuclear qudit system.
The combination of noise-resilient nuclear degrees of freedom and large reduction of nutation time induced by electron-nuclear mixing  %large hyperfine interaction
enables coherent manipulation of this qudit by radio-frequency pulses. Moreover, the multi-level structure of the qudit is exploited to encode and operate a qubit with embedded basic quantum error correction.
\end{abstract}

\begin{figure}[h]
	\centering
	\includegraphics[width=0.48\textwidth]{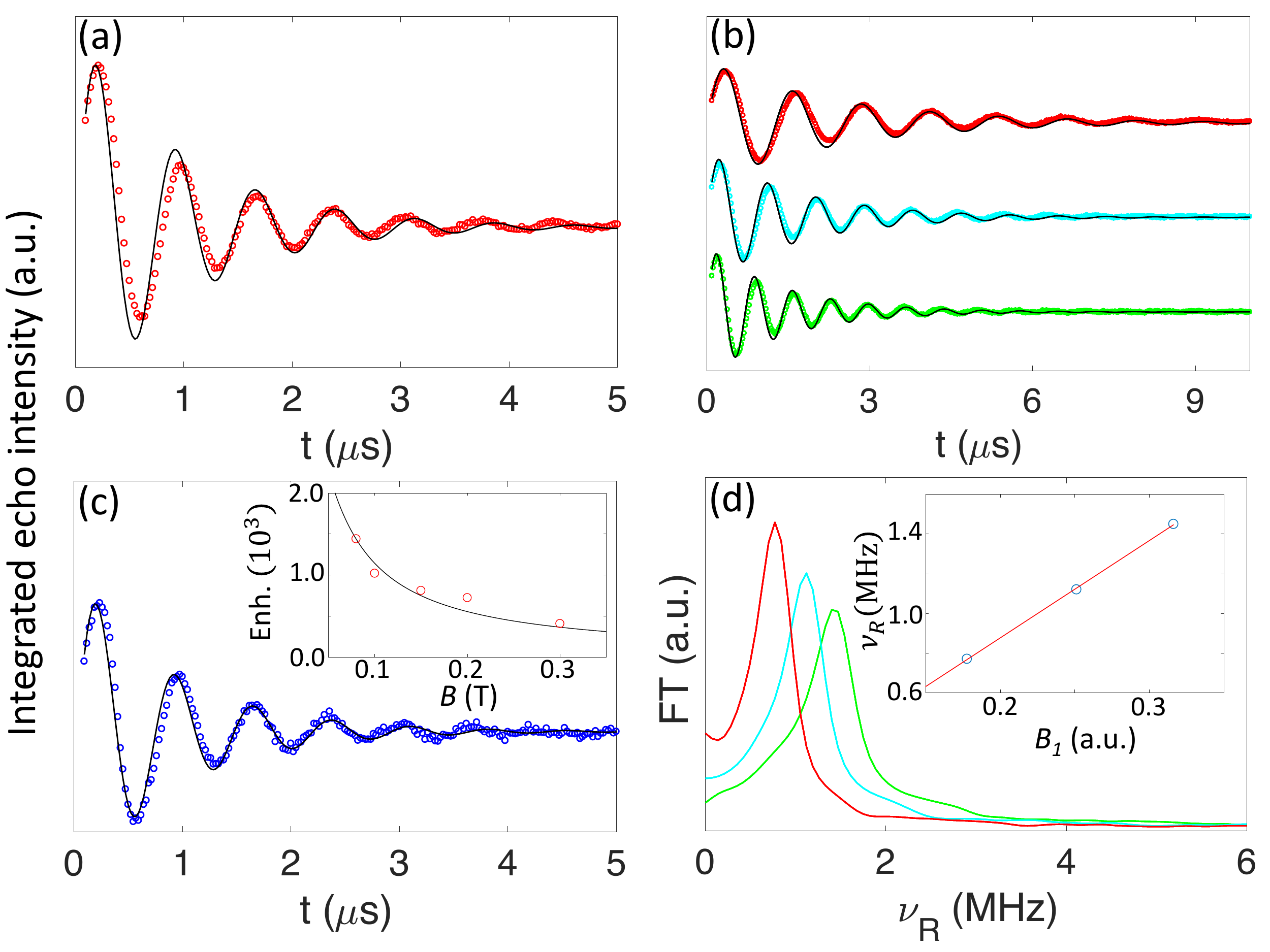}
	\caption{(a-c) Rabi oscillations for transitions F1, F3, F4 necessary to operate the error-protected qubit. Pulses of different duration are used to induce Rabi oscillations, detected by a final $\pi$-pulse for refocusing. Lines are fits with the damped functions $\propto e^{-t/\tau} \text{sin} [2\pi \nu_R t ]$. Different radiofrequency powers are reported for F3 (b). Inset of (c): measured (dots) and calculated (line) enhancement of $\nu_R$ as a function of $B$. (d) Fourier transform, evidencing the  monochromatic character of the oscillations in (b). Inset: linear scaling of $\nu_R$ with $B_1$.}
	\label{fig:Rabi}
\end{figure}
\begin{figure}[h]
	\centering
	\includegraphics[width=0.48\textwidth]{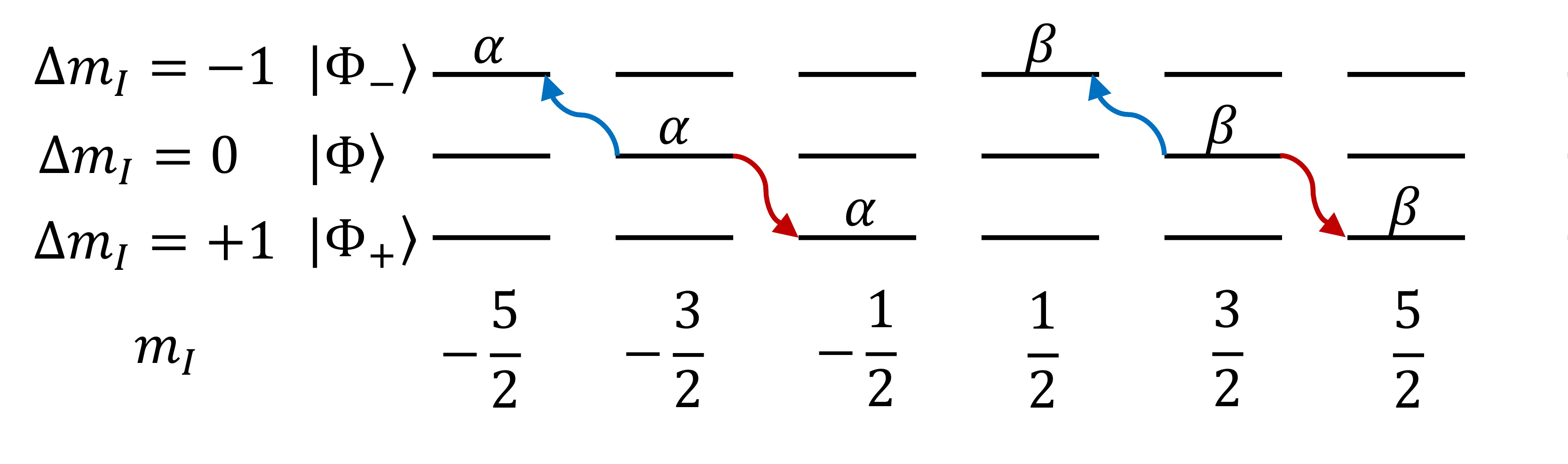}
	\caption{Encoding of a qubit protected from amplitude shift errors in the nuclear levels of {\bf 1}, corresponding to the $m_S=-1/2$ manifold. 
		Possible amplitude shift errors are indicated by arrows.}
	\label{fig:qudit}
\end{figure}

The realization of quantum computers is one of the hottest topics of current research. The elementary units of these devices are quantum two-level systems (qubits), that can be realized using a variety of physical objects \cite{RMPNori}, including tailor-made molecules hosting electron spins. In particular, molecular transition-metal spin qubits with remarkably long coherence times were recently synthesized \cite{PRLArdavan,FreedmanDeco,Bader,Freedman2015,jacsVO,Freedman2016VO,jacsVOc,ChemSciVO}, thus enabling the demonstration of 
single-qubit gates by pulsed electron paramagnetic resonance (EPR) \cite{V15,Freedman2014,Freedman2016Cr,jacsVOb}. 
Coordination chemistry was exploited to link molecular qubits \cite{NatNano,entang}, to obtain switchable couplings between them \cite{SciRep,modules,Chem} and to experimentally demonstrate two-qubit gates \cite{TakuiCNOT,Ardavan}.
Ln complexes are a rather unexplored but very interesting class of molecular qubits
\cite{Hill,PRLLuis,JacsLuis,Stergios2016,luminescence}. 
Among the few examples, a very promising qubit is [Yb(trensal)] of the [Ln(trensal)] series\cite{Kanesato,Flanagan,ChemCommtrensal}, in which coherent Rabi oscillations of the electronic spin have been recently reported \cite{Stergios2016}.  \\
Compared to qubits, quantum systems with $d>2$ levels, called qudits, provide additional resources that can enhance the power of quantum logic  \cite{srep2017, PRLBullock,NatPhys2008,PRBLuisGd,PLA2015,Science2009,Chau,Pirandola,Crypt,NatPhys2011}.
For instance, the four-levels nuclear spin of a Tb ion was recently exploited to implement Grover's quantum algorithm \cite{GroverWW2018}.
Magnetic nuclei were identified as a potential resource for molecule-based quantum information processing \cite{Takui,ScienceWW,TakuiPCCP,GroverWW2018,Vanadili2018}, thanks to their isolation from the environment, which however leads to long manipulation times \cite{Jones}.
This limitation can be overcome in magnetic ions hosting magnetic nuclei, where the
electronic degrees of freedom can 
significantly speed-up quantum gates, thanks to the hyperfine coupling\cite{Mehring,Cory,LaflammePRL}.
Moreover, electrons can be exploited to obtain an effective switchable coupling between nuclear qubits\cite{Vanadili2018}. \\
Here we show that $^{173}$Yb isotope of [Yb(trensal)] \cite{InorgChem} is a prototypical coupled electronic qubit-nuclear qudit system. Indeed, its $I=5/2$ nucleus provides a six-level qudit that can be rapidly manipulated thanks to its strong hyperfine interaction with the electronic effective $S=1/2$. We demonstrate coherent control of the nuclear-spin degrees of freedom of [$^{173}$Yb(trensal)] by nuclear magnetic resonance (NMR). 
We use NMR to accurately characterize the system and spin-echo pulse sequences to determine  spin-lattice relaxation and spin-coherence times. These latter are significantly enhanced by implementing Carr-Purcell-Meibum-Gill (CPMG) sequences.
We observe fast Rabi oscillations of the nuclear magnetization, demonstrating our capability to coherently manipulate the $^{173}$Yb qudit. The multi-level structure of the qudit can be exploited to encode an error-protected logical qubit.
In particular, we show that a minimal code\cite{Pirandola} protecting against amplitude or phase shift errors can be implemented.\\
Measurements were performed at $T = 1.4$ K on a single-crystal of [Yb(trensal)], doped  at 2\% (as determined by ICP-MS) into the isostructural diamagnetic host [Lu(trensal)] ({\bf 1}), to reduce electronic dipolar interactions. {\bf 1} crystallizes in the $P\bar{3}c1$ space group with the Yb$^{3+}$ ion and the apical tertiary amine nitrogen atom lying on the crystallographic C$_{3}$ axis\cite{Stergios2016}. 
\begin{figure}[h]
	\centering
	\includegraphics[width=0.45\textwidth]{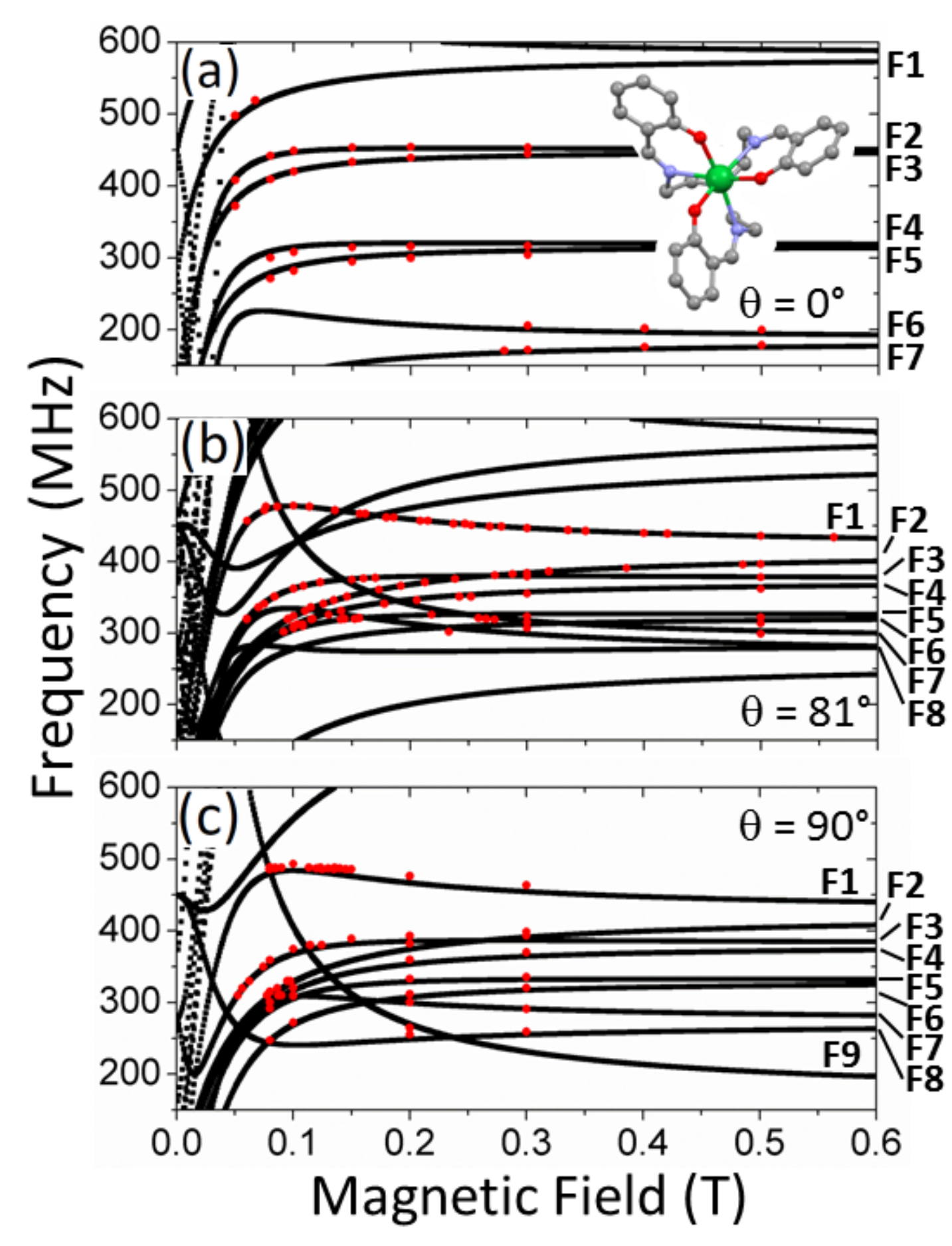}
	\caption{(a-c) Measured (red circles) and calculated (black) frequencies of relevant transitions for different orientations of ${\bf B}$ with respect to the crystal C$_3$ axis. Transitions are labeled in order of increasing frequency at $0.6$ T. Inset: structure of {\bf 1}. }
	\label{fig:levels}
\end{figure}
The electronic ground doublet is well isolated in energy from excited crystal-field states\cite{InorgChem}, hence we model {\bf 1} as an effective spin $S=1/2$ coupled to a nuclear spin $I$. The Yb natural composition encompasses several isotopes, of which only the $^{171}$Yb and $^{173}$Yb have a non-zero nuclear spin. We focus on the $^{173}$Yb isotope ($I$ = 5/2), that can encode a $d=6$ qudit. Transitions due to the $^{171}$Yb isotope occur at very different frequencies and do not influence our study.\\
The system is described by the effective Hamiltonian:
\begin{eqnarray}\nonumber
H &=& A_\parallel S_z I_z + A_\perp \left( S_x I_x + S_y I_y\right) + p I_z^2 \\
&+&  \mu_B \textbf{S} \cdot \textbf{g} \cdot \textbf{B} + \mu_N g_I \textbf{I} \cdot \textbf{B},
\label{spinHam}
\end{eqnarray}
where the first two terms describe the hyperfine coupling, the third is the nuclear quadrupolar interaction and the last two represent the electronic and nuclear Zeeman terms. We use $g_x=g_y=2.9$ and $g_z=4.3$ as determined by EPR\cite{InorgChem}.
The parameters of (\ref{spinHam}) have been determined from the position of the peaks in the NMR spectra (see SI). The full set of the observed peak frequencies is reported in Figure \ref{fig:levels}, for three different orientations $\theta=0^\circ, 81^\circ, 90^\circ$ of ${\bf B}$ with respect to the C$_3$ axis.
Experimental frequencies (red dots) are well reproduced by the model (black curves), with $A_\parallel = -0.02993 ~\text{cm}^{-1}$, $A_\perp = -0.0205 ~\text{cm}^{-1}$ (close to those determined by EPR \cite{InorgChem}) \footnote[8]{The sign of the hyperfine coefficients was taken from Ref. \citenum{InorgChem}.}, $g_I=-0.2592$ \cite{Stone} and $p = -0.0022 ~\text{cm}^{-1}$, which is reasonable for $^{173}$Yb \cite{AbragamNMR}.
We note that the previously undetermined quadrupolar term $p$ is essential to reproduce NMR data. \\
The level diagrams obtained from diagonalization of (\ref{spinHam}) are reported in the SI. A large anti-crossing at small fields is due to the entanglement of electronic and nuclear spins, induced by the component of the hyperfine tensor perpendicular to ${\bf B}$. If we focus on $\theta=0^\circ$, an expression for the eigenstates can be obtained for $g_z \mu_B B>>|A_\perp|$ (satisfied if $B \gtrsim$ 0.1 T).
To first order in perturbation theory (neglecting, for simplicity, nuclear quadrupolar and Zeeman interactions), we get 
\begin{equation}
|\psi_{m_S, m_I}\rangle \! \propto \!  |m_S, m_I\rangle \pm \frac{\alpha A_\perp |m_S\mp1, m_I\pm1\rangle}{g_z \mu_B B + (m_I \pm 1/2) A_\parallel}  ,
\label{mixing}
\end{equation}
with $m_S, m_I$ the eigenvalues of $S_z$ and $I_z$. The $\pm$ signs depend on the value of $m_S$, while
%$\alpha= \langle m_S\pm 1, m_I \mp 1 | S_x I_x+S_y I_y  |m_S, m_I\rangle = \langle m_I \mp 1 | I_x  |m_I\rangle$
$\alpha A_\perp$ is the matrix element of the transverse hyperfine interaction.
The matrix element of NMR transitions $|\psi_{m_S, m_I}\rangle \rightarrow |\psi_{m_S, m_I \pm1}\rangle $  is approximately proportional to
$g_I\mu_N \pm \frac{A_\perp}{2B}$, where the first term corresponds to a pure nuclear excitation and the second arises from electron-nuclear mixing.
It is worth noting that even a rather small mixing of the wave-function ($\sim 0.08$ at 0.15 T), induces a large enhancement of the matrix element (with respect to the case $A_{\perp}=0$), since $\frac{A_\perp}{2B} \sim 10^3~g_I\mu_N$. Therefore, manipulations of the nuclear qudit are actually very fast in {\bf 1} (see below). \\
\begin{figure}[h]
	\centering
	\includegraphics[width=0.48\textwidth]{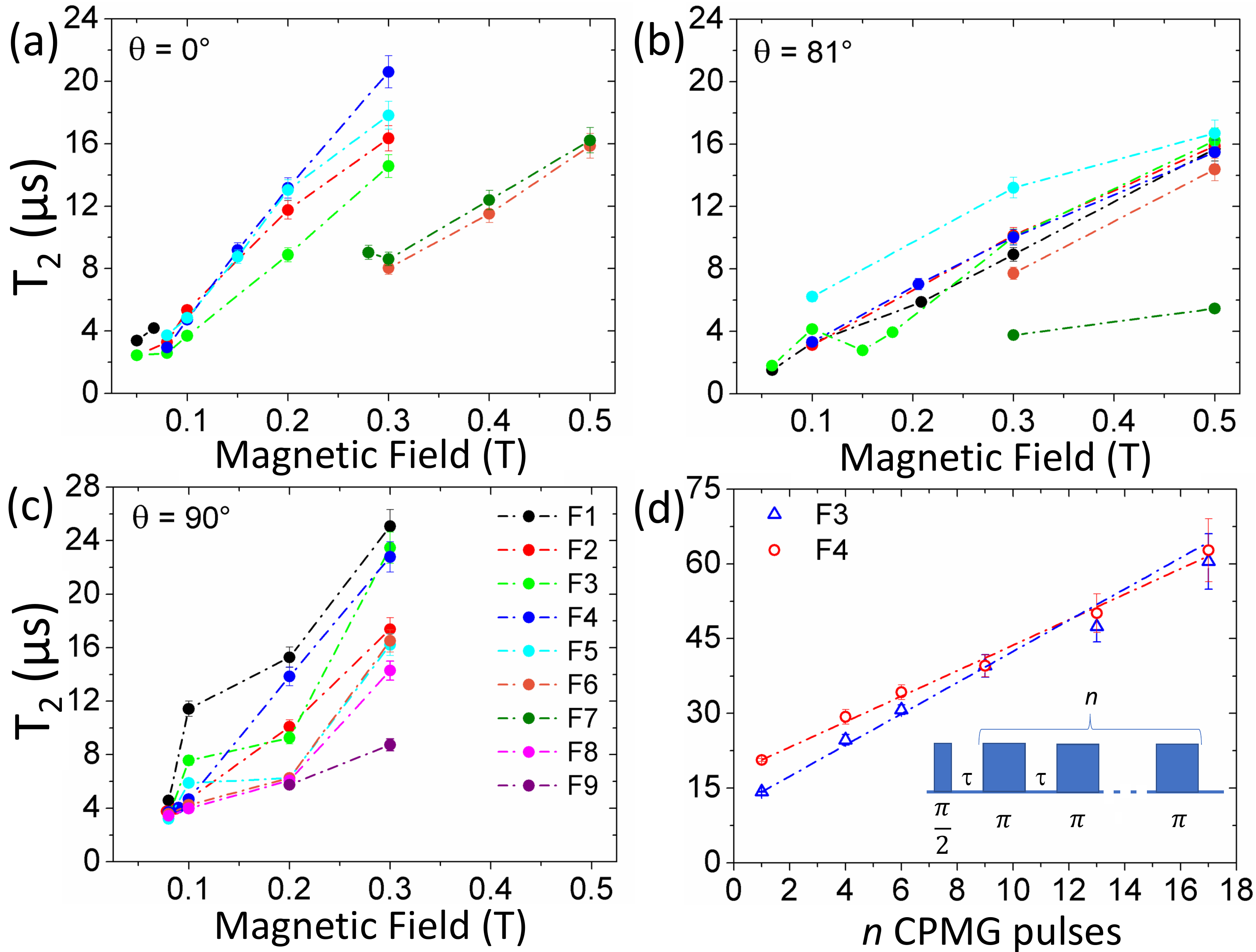}
	\caption{(a-c) Coherence time $T_2$ measured with Hahn echo sequence for transitions indicated in Figure \ref{fig:levels} and different orientations of the field. (d) Enhancement of spin coherence with the number $n$ of refocusing $\pi$ pulses in CPMG sequences (inset). Lines are guides for the eye. }
	\label{fig:T2}
\end{figure}
To investigate the coherence times of the nuclear qudit, Hahn echo was detected for all transitions indicated in Figure \ref{fig:levels} \cite{Hahn}. 
%by applying the $\frac{\pi}{2}-\tau-\frac{\pi}{2}-\tau$ pulse sequence \cite{Hahn}. 
The decay of the nuclear magnetization is fitted by a single-exponential function (see SI). The resulting coherence time $T_2$ is reported in Figure \ref{fig:T2}-(a-c) as a function of $B$, with $\theta = 0^\circ, 81^\circ, 90^\circ$. 
We note that the observed $T_2$ increases with the magnetic field. 
Indeed, the decay of the Yb nuclear coherence is caused either by the interaction with the neighboring nuclei or by dipolar couplings with other electronic spins,
%residual dipole-dipole couplings
largely reduced in our diluted sample.
These mechanisms are mediated by the electron-nuclear mixing (Eq. \ref{mixing}), which decreases with $B$, explaining the observed field dependence of $T_2$. Moreover,
the electronic polarization increases with $B$, thus reducing spin flops \cite{jpcl}. \\
The spin coherence time can be further increased by employing a CPMG pulse sequence\cite{Slichter}, i.e. a concatenation of $n$ refocusing pulses to dynamically decouple the system from the environment \cite{ScienceDD}.
This results in a clear enhancement of the observed coherence time with $n$ \cite{Lloyd,Uhrig,PRLscalingT2}, as shown in Figure \ref{fig:T2}-(d) for transitions F3  and F4 ($\theta=0^\circ$). Details are provided in the SI.\\
The ability to generate arbitrary coherent superposition states is demonstrated by transient nutation experiments. 
Figure \ref{fig:Rabi}-(a-c) reports measured Rabi oscillations, showing that few hundreds of ns are sufficient to implement $\pi$ rotations. 
The loss in coherence time induced by the sizable electron-nuclear mixing (at $B \lesssim 0.3$ T) is compensated by the enhancement in the Rabi frequency ($\nu_R$), which makes these gates particularly fast. Indeed, $\nu_R$ is $10^3$ times larger compared to the case $A_{\perp}=0$ (inset of Fig. \ref{fig:Rabi}-(c)), in very good agreement with calculations based on (\ref{spinHam}).
%As discussed above, the sizable electron-nuclear mixing at $B \lesssim 0.3$ T, while reducing coherence times, makes the implementation of these gates particularly fast. Indeed, the Rabi frequency is here enhanced by almost $10^3$ with respect to the case $A_{\perp}=0$ (see e.g., inset of Fig. \ref{fig:Rabi}-(c)), in very good agreement with calculations based on (\ref{spinHam}).
The linear scaling of $\nu_R$ with the oscillating field amplitude $B_1$ [Figure \ref{fig:Rabi}-(d)] confirms that these are Rabi manipulations. Their damping increases with the applied power and is mainly due to inhomogeneities of $B_1$ (SI and Ref. \citenum{Chiorescu}).  \\
The presence of an effective non-axial electronic ground doublet, the
observed values of $T_2$, the enhancement in $\nu_R$ ensured by the large $A_\perp$ and the fact that transitions are well resolved (Fig. \ref{fig:levels}) make {\bf 1} a promising coupled electronic qubit-nuclear qudit system. 
Here we show that the multi-level structure ($d=6$) of the $^{173}$Yb nucleus can be used to encode a qubit protected against amplitude or phase shift errors \cite{Pirandola}, whereas the electronic qubit provides a fast ancilla to detect errors and help efficient gates implementation. We start by focusing on amplitude shifts, i.e. unwanted $\Delta m_I  = \pm 1$ transitions. By encoding the logical qubit in the generic state $\alpha \vert 0\rangle + \beta \vert 1 \rangle $ in $\vert \Phi \rangle = \alpha \vert \psi_{m_S=-1/2, m_I=-3/2} \rangle + \beta \vert \psi_{m_S=-1/2, m_I=3/2} \rangle$ (with ${\bf B}$ along $z$), the effect of an amplitude shift is to induce a transition to a state outside the computational basis. Hence, the error can be detected and then corrected \cite{Pirandola}. In the case of a $\Delta m_I = 1$ shift $\vert \Phi \rangle \rightarrow \vert \Phi_+ \rangle = \alpha \vert \psi_{-1/2, -1/2} \rangle + \beta \vert \psi_{-1/2, 5/2} \rangle$. This state conserves the superposition and has no overlap with $\vert \Phi \rangle$ (see Figure \ref{fig:qudit}). Thus, by two simultaneous and fast microwave $\pi$ pulses resonant with the $\vert \psi_{-1/2, -1/2} \rangle \rightarrow \vert \psi_{1/2, -1/2} \rangle$ and $\vert \psi_{-1/2, 5/2} \rangle \rightarrow \vert \psi_{1/2, 5/2} \rangle$ gaps, 
(echo detected in Ref. \citenum{Stergios2016}), it is possible to rotate the electronic spin only if the molecule is in $\vert \Phi_+ \rangle$. Therefore, a measure of the electronic spin after the pulses detects and identifies the error, without affecting the qubit state (i.e., without collapsing the nuclear wavefunction). After restoring the ancilla in $m_S=-1/2$ (by repeating the microwave pulses), the error can be corrected by radio-frequency pulses. An analogous procedure detects and corrects $\Delta m_I = -1$ errors leading to $ \vert \Phi_- \rangle = \alpha \vert \psi_{-1/2, -5/2} \rangle + \beta \vert \psi_{-1/2, 1/2} \rangle$.
We stress that the minimal number of qudit levels to implement this code is exactly 6 as in the present case ($I=5/2$), because it enables $\vert \Phi \rangle$, $ \vert \Phi _+\rangle$ and $ \vert \Phi _-\rangle$ not to overlap. This quantum error correction code requires less levels than the textbook 3-qubit repetition code \cite{Nielsen}.\\
An important point is to implement a universal set of single-qubit gates with the present error-protected encoding. Rotations around $x$ and $y$ of $\vert \Phi \rangle = \alpha \vert \psi_{-1/2, -3/2} \rangle + \beta \vert \psi_{-1/2, 3/2} \rangle$ can be implemented by a sequence of radiofrequency pulses resonant with the $-3/2 \rightarrow -1/2$, $-1/2 \rightarrow 1/2$ and $1/2 \rightarrow 3/2$ gaps (see SI), i.e. %F1, F3 and F4 in Figure \ref{fig:levels}. 
exactly the fast Rabi manipulations reported in Fig. \ref{fig:Rabi}. Hence, rotations around $x$ and $y$ of the encoded-qubit state can be implemented.
Furthermore, rotations about $z$ of $\vert \Phi \rangle$ can be easily obtained by a single microwave $2\pi$ pulse semiresonant with the $\vert \psi_{-1/2, 3/2} \rangle \rightarrow \vert \psi_{1/2, 3/2} \rangle $ gap (see SI). We finally note that the complementary phase-shift code could be implemented in the same way by encoding in the conjugate basis \cite{Pirandola}.\\
In conclusion, we have shown that the [$^{173}$Yb(trensal)] molecule behaves as a nuclear qudit coupled to an electronic qubit. We have experimentally demonstrated that the nuclear qudit can be rapidly and coherently manipulated, thanks to the combination of a strong hyperfine interaction and long coherence times, which can be further enhanced by CPMG sequences. Thanks to its nuclear qudit, [$^{173}$Yb(trensal)] can encode a qubit with embedded basic quantum error correction. These results open new perspectives in the relatively new field of $f$-electrons molecular qubits. In particular, the next step would be to investigate a dimer of interacting Yb molecular units enabling an error-resilient two-qudit gate.

\begin{acknowledgement}

The authors acknowledge financial support from PRIN Project 2015 HYFSRT of the MIUR (Italy) and from the European Project SUMO of the call QuantERA. S.P. acknowledges VILLUM FONDEN for research grant 13376.\\
This document is the Accepted Manuscript version of a Published Work that appeared in final form in J. Am. Chem. Soc., copyright \textcopyright American Chemical Society after peer review and technical editing by the publisher.
To access the final edited and published work please use the following DOI:10.1021/jacs.8b05934

\end{acknowledgement}

\begin{suppinfo}
\begin{itemize}
  \item Synthesis and characterization, details concerning NMR measurements, spin Hamiltonian calculations, details on Quantum Error Correction.
\end{itemize}

\end{suppinfo}

\bibliography{Ybtrensal}

\end{document}